\documentclass[12pt]{article}
\usepackage{psfig,epsf}
\usepackage{a4wide,amssymb,graphicx}
\usepackage{cite}

\def\MeV{\textrm{ MeV}}

\newcommand{\beq}{\begin{equation}}
\newcommand{\eeq}[1]{\label{#1} \end{equation}}

\newcommand{\bed}{\begin{displaymath}}
\newcommand{\eed}{\end{displaymath}}
\def\bea{\begin{eqnarray}}
\def\eea{\end{eqnarray}}
\begin{document}

\title{ Testing the nature of the $\Lambda (1520)$ resonance
through photoproduction}

\maketitle

\begin{center}
{\large{ L.~Roca$^{1,2}$, E.~Oset$^{1,2}$ and H. Toki$^1$}}

\vspace{.5cm}
{\it $^1$ Research Center for Nuclear Physics (RCNP), Osaka University,\\
Ibaraki, Osaka 567-0047, Japan}\\
{\it $^2$ Departamento de F\'{\i}sica Te\'orica and IFIC
 Centro Mixto Universidad de Valencia-CSIC\\
 Institutos de Investigaci\'on de Paterna, Apdo. correos 22085,
 46071, Valencia, Spain}

\end{center}

\begin{abstract}  Recent studies within the framework of
chiral unitary theory produce the $\Lambda (1520)$ resonance,
among several others, as a dynamically generated resonance
from the interaction of the baryon decuplet and the meson
octet. The $\Lambda (1520)$ qualifies as a quasibound state
of $\pi \Sigma(1385)$ and this has repercussion in some
observable quantities. In the present work we show that the
$\gamma p\to K^+K^-p$ reaction has a sizeable cross section
for invariant $K^-p$ masses above the $\Lambda(1520)$ mass.
On the other hand, we also find that the 
$\gamma p\to K^+\pi\Sigma(1385)$ reaction has a sizeable cross section in
that energy region as a consequence of the strong coupling of
the  $\Lambda (1520)$ to  $\pi\Sigma(1385)$, and then we make
predictions for the ratio of this cross section to that of
the $\gamma p\to K^+ K^-p $ reaction.

\end{abstract}

The introduction of unitary techniques in the study of meson
 baryon interaction with chiral Lagrangians has allowed 
to show that many of the low lying baryonic resonances
 are dynamically generated in the implicit process of
 multiple scattering, or equivalently, they qualify as
 quasibound meson baryon states.  Early studies in this
 direction pointed at the $\Lambda(1405)$ \cite{norbert} and
 the $N^*(1535)$ \cite{siegel} as dynamically generated
 resonances.  The unitary methods to deal with the meson
 baryon interaction within the chiral framework have been
 made more systematic and a variety of them are now
 available, leading to remarkably similar results
 \cite{angels,jose,ulf,bennhold,bennhold2,jido,nam,lutz,juan}. 
 One of the  consequences of these studies is that the
 interaction of the baryon octet and meson octet leads to
two octets and one singlet of dynamically
 generated resonances with $J^P=1/2^-$ \cite{juan,jido},  to
 which the  $\Lambda(1405)$ and the $N^*(1535)$ belong.

     An interesting follow up of these  developments has been
recently done in \cite{lutz} with the interaction of the
baryon decuplet with the meson octet. Also in this case a
number of $J^P=3/2^-$ resonances are dynamically generated,
which are easily identified with known resonances in the
 PDG\cite{pdg},
and some other ones are predicted. These results have been
confirmed in another study where  poles and residues in the
complex plane are searched for \cite{sarkar}, and which
allows one to get the couplings of the resonance to the
different coupled channels. Among these states there is one
suggested in \cite{lutz}, and studied in detail in \cite{kdel},
and is formed from $\Delta K $ interaction with quantum
numbers $S=1$, $I=1$, and is an exotic baryon impossible to
construct with three constituent quarks, hence, a resonance as
exotic as the $\Theta ^+$ discovered  in Spring8/Osaka
\cite{nakano} (see also Proceedings of the pentaquark04
Workshop for an update of the experimental and theoretical
status of this issue  \cite{pentawork} ).

   One of the side effects of the study of the  $\Theta ^+$
state is the test for the production of standard resonances,
which is conducted both as a proof that the methods used to
identify resonances do indeed work, and also to determine
regions of phase space which are ideal to reduce backgrounds
and, hence, see a clearer signal of the $\Theta ^+$ state. 
One of the resonances thoroughly studied at Spring8/Osaka is
the $\Lambda(1520)$ in the $\gamma p\to K^+ K^-p$ reaction
using photons of $2.0$ to $2.4\textrm{ GeV}$ \cite{nakano2}.
The $\Lambda(1520)$ peak is clearly seen in the $K^+$ missing
mass spectrum, which also shows a sizeable background at
energies above the peak, even when the  background from kaons
coming from $\phi$ decay is eliminated. Another experimental
work on this reaction was done in \cite{dares} using photons
from $2.8-4.8\textrm{ GeV}$, and similar features as in
\cite{nakano2} were observed.  We shall see that the
interaction of coupled channels which leads to the
$\Lambda(1520)$ pole, together with the D-wave character of
the $\Lambda(1520)$ resonance in its $\bar{K}N$ decay, can
explain this background. 

  The $\Lambda(1520)$  is one of the dynamically generated
resonances in \cite{lutz,sarkar}.  It appears from the
interaction of the coupled channels  $\pi \Sigma(1385)$ and
$K \Xi(1530)$, mostly the first one. In addition, the
$\Lambda(1520)$ mass is just  about 5 MeV below the $\pi
\Sigma(1385)$ threshold.  All these things make the 
$\Lambda(1520)$ qualify as a quasibound $\pi \Sigma(1385)$
state.  However, the small branching ratio of the
$\Lambda(1520)$ to $\pi \Sigma(1385)$ of about 4 percent in
the only experiment available \cite{pdg,chan} (about 10
percent  assuming, as done in the PDG, that the $\Lambda \pi
\pi$ channel is mostly   $\pi \Sigma(1385)$) does not seem to
indicate such a large coupling of the $\Lambda(1520)$ 
resonance to the $\pi \Sigma(1385)$ state.  Actually, for the
nominal values of the masses, the decay of the
$\Lambda(1520)$ into this channel is forbidden since 1520 MeV
is about 5 MeV below threshold of  $\pi \Sigma(1385)$. 
Hence, it is the width of the $\Lambda(1520)$ and the 
$\Sigma(1385)$, which are both rather narrow, what makes the
decay possible, however, relatively small.  Also, when the
channels are so close to threshold  the branching ratio  to
these channels is always partially a matter of  choice since
it depends on the energy cuts one is taking. A cleaner
observable to find the coupling of the $\Lambda(1520)$
resonance to $\pi \Sigma(1385)$ is hence called for.
A consequence of the nature of the $\Lambda(1520)$ as an
approximately quasibound state of $\pi\Sigma(1385)$ is a
relatively large coupling of the resonance to this channel,
which the chiral unitary approach provides \cite{sarkar}. One
of the issues we address in this paper is how this coupling
could be determined experimentally, which could shed light on
the nature of the $\Lambda(1520)$ resonance. For this purpose
we suggest the measurement of the ratio of cross sections for
the reactions
\begin{eqnarray}
\nonumber \gamma p &\to& K^+ K^- p\\ 
\gamma p &\to& K^+ \pi \Sigma(1385),
\label{eq:reac}
\end{eqnarray}  
and we evaluate this ratio within the framework of the chiral
unitary approach.

Preliminary results for the first reaction of
Eq.~(\ref{eq:reac}) have been obtained at Spring8/Osaka
\cite{nakano2}, and experimental results are also available in
\cite{dares}. One observes there a clear peak of the
$\Lambda(1520)$ in the $K^-p$ invariant mass distribution on
top of a moderate background at masses beyond the
$\Lambda(1520)$ peak. Another aim of the
present work is to show that such a background appears
naturally within the chiral unitary approach, which provides
scattering amplitudes and not just poles of resonances.

\begin{figure*}[htb]
\begin{center}
\includegraphics[height=2cm]{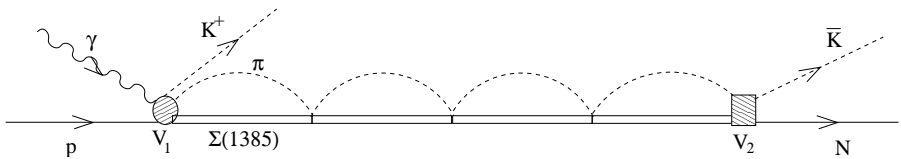}
\end{center}
\caption{Schematic representation of the mechanism for the 
$\gamma p\to K^+K^-p$ reaction mediated by the
$\Lambda(1520)$, which is generated through multiple
scattering of $\pi\Sigma(1385)$ implicit in the diagram.} 
\label{fig:unit}
\end{figure*}

Since the $\Lambda(1520)$ is dynamically generated from the
$\pi \Sigma(1385)$ (and to a much lesser extend the
$K\Xi(1530)$ coupled channel \cite{sarkar}), the microscopic
description of the $\gamma p\to K^+K^-p$ process would be
given diagrammatically by the mechanism in
Fig.~\ref{fig:unit}, in analogy whith what was done in
\cite{nacher} for the photoproduction of the $\Lambda(1405)$.

We shall try to make our results the least model dependent
possible, hence we avoid making an explicit model for the
$\gamma p \to K^+ \pi \Sigma(1385)$ amplitude at tree level
($V_1$) and similarly we also avoid making a model for the
final $\pi \Sigma(1385)$ state decaying into $\bar{K} N$. 
The only important thing to keep in mind is that the
transition from the $\Lambda(1520)$ to $\bar{K} N$  proceeds
in D-wave and this implies a factor $q^2$ in the transition
amplitude, with $q$ the $\bar{K}$ momentum in the $\bar{K} N$
center of mass frame. The mechanism also involves the $t_{\pi
\Sigma^* \to \pi \Sigma^*}$ amplitude which implicitly
contains the $\Lambda(1520)$ pole. For the rest of the
amplitudes we shall assume a smooth energy dependence,
although this assumption will be unnecessary when we study
the ratio of the cross sections of the two reactions in 
Eq.~(\ref{eq:reac}), where the same vertex appears in both
cases. 

The amplitude for the $\gamma p \to K^+ K^- p$
reaction will be given by

\begin{equation}
 t=\widetilde{V}_1 t_{\pi \Sigma^* \to \pi
\Sigma^*} \widetilde{V}_2 q^2,
\label{eq:t}
\end{equation} 

 \noindent where
$\widetilde{V}_1$, $\widetilde{V}_2$,  are the vertices of
Fig.~\ref{fig:unit} including, respectively,
 the first and last loop
functions of Fig.~\ref{fig:unit}, involving the
$\pi$ and $\Sigma(1385)$ propagators. We shall assume 
$\widetilde{V}_1$ and $\widetilde{V}_2$
 smooth compared with the  $t_{\pi \Sigma^* \to \pi
\Sigma^*}$ amplitude which incorporates the  $\Lambda(1520)$
structure. The details on how to evaluate this latter
amplitude can be seen in \cite{sarkar}. 
In Eq.~(\ref{eq:t}), $t_{\pi \Sigma^* \to \pi \Sigma^*}$
stands for the scattering amplitude of 
$\pi \Sigma(1385) \to \pi \Sigma(1385)$ in isospin $0$ 
which is obtained in
\cite{sarkar} within the framework of a chiral unitary
approach using the $N/D$ unitarization method in coupled
channels \cite{jose} (or equivalently the Bethe-Salpeter
equation \cite{angels}) and contains the resummation of the
Dyson series involving $\pi\Sigma(1385)$ and $\pi\Xi(1530)$
loops.

The $K^- p$ invariant mass distribution for the reaction
$\gamma p\to K^+(p_3)K^-(p_2)p(p_1)$ is given by

\begin{equation}
\frac{d\sigma}{dM_{12}}= D \int_{m_1}^{M-E_3-m_2}
 dE_1 M_{12} \Theta(1-A^2)
\overline{\sum}|t|^2,
\end{equation}

\noindent
where $D$ is supposed to be a constant and $A$ stands for
the cosinus of the $\vec{p}_1$ and $\vec{p}_3$
angle which is fixed by the other variables and given by
\begin{equation}
A=\cos\theta_{13}=\frac{(M_{12}-E_1-E_3)^2-m_2^2-\vec p\,^2_1
-\vec p\,^2_3}
{2p_1p_3}.
\label{eq:A}
\end{equation}

  Application of the results of \cite{sarkar} to the present
work requires some fine tunning which we describe here.  In
\cite{sarkar} the position of the $\Lambda(1520)$ appears
around 1560 MeV when using a global subtraction constant in
the dispersion relation formula of \cite{sarkar}
to reproduce the bulk of the
$3/2^-$ resonances, and the width is larger than the nominal
one. This is because the $\pi\Sigma^(1385)$ channel is open
at these energies.  We can do fine tunning, changing the
subtraction constant from $a=-2$ to $a=-2.72$, which brings the
position of the resonance down to 1520 MeV and zero width,
since it is below the $\pi\Sigma(1385)$ threshold and we are
ignoring the $\Sigma(1385)$ width and the $\bar{K} N$ and
$\Sigma N$ decay channels.

We calculate the coupling, $g$, of the $\Lambda(1520)$ resonance
to the  $\pi \Sigma(1385)$ channel by means of the residue of
the  $\pi \Sigma(1385) \to \pi \Sigma(1385)~ I=0$ amplitude,
which close to the pole behaves as
\begin{equation}
   \frac{g^2}{z-z_R}.
\label{eq:pole}
\end{equation}   
We obtain $|g|=1.21$ with that procedure and assume a
conservative error for $g$ of $20$ percent.
Half of this uncertainty comes from varying the pole position of
the $\Lambda(1520)$ within the experimental errors in the mass.
The rest of the error would come by assuming an uncertainty of
abot $7\textrm{ MeV}$ in the $\Sigma(1385)$ mass to partly
account for its width and from 
our neglect of the $\bar K N$ and
$\pi\Sigma$ channels in the build of the $\Lambda(1520)$
resonance.  
As mentioned above, we obtain a $\Lambda(1520)$ pole in the real
axis with this prescription, and we have used the procedure to
obtain the coupling $g$ which is rather stable with respect to
small variations of the parameters.
 
On the other hand,  in order to have a shape for the  
$\Lambda(1520)$ excitation similar to the one found in
\cite{dares}, we  change the subtraction constant
to $a\simeq -2.51$ to $-2.54$ and simultaneously the mass of
the  $\Sigma(1385)$ by about $7\MeV$ (to simulate
contributions coming from the consideration of the
$\Sigma(1385)$ width). By means of this,
 we obtain a finite  width, which
allows for a realistic distribution of the strength of the
resonance.  

\begin{figure*}[htb]
\begin{center}
\includegraphics[height=9cm]{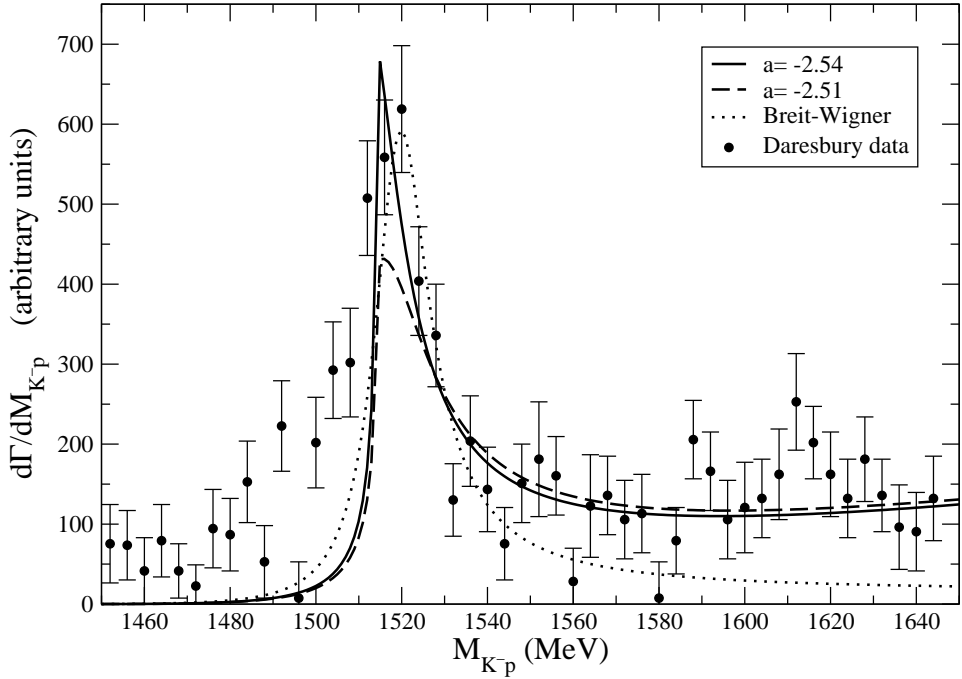}
\end{center}
\caption{$K^-p$ invariant mass distribution of the $\gamma p\to
K^+K^- p$ reaction with photons in the range
$E_\gamma=2.8-4.8\textrm{ GeV}$. (The theoretical curves are an
average within this range).
Experimental results from \cite{dares}.} 
\label{fig:dares}
\end{figure*}

In Fig.~\ref{fig:dares} we show the experimental results for
$d\sigma/dM_{K^-p}$ in the $\gamma p\to K^+K^-p$ reaction
from Daresbury \cite{dares}, together with results of our
model for the two different values of the subtraction
constants which have been fine tuned to get an approximate
agreement with the experimental data from the $\Lambda(1520)$
peak onward. We have not made any attempt to reproduce the
data below the $\Lambda(1520)$ peak since the deficiencies of
our model (not including the $\bar K N$ and $\pi\Sigma$
channels, the only ones open below the $\pi\Sigma(1385)$
threshold)
 do not allow for a realistic description of the data
in that region. However, our model contains the coupling of
the $\Lambda(1520)$ to $\pi\Sigma(1385)$ which is largely
dominant, and as soon as there is phase space for
$\pi\Sigma(1385)$ it becomes mostly responsible for the
strength of the distribution. It is interesting to note that
the fine tuning of the subtraction constant changes moderately
the strength at the peak but barely changes the strength in
the region of $1550-1650\MeV$.
The comparatively large strength of the distribution at
energies higher than the peak is due to the large
$\pi\Sigma(1385)\to\pi\Sigma(1385)$ amplitude in this region,
together with the $q^2$ character of the $D-$wave transition
amplitude $\pi\Sigma(1385)\to\bar{K}N$. In order to give an idea
of the relative size of this strength in this region, we show
in the same figure the distribution produced by a naive
Breit-Wigner resonance around the peak (including also the
$q^2$ factor in the amplitude). We see a sizeable difference
which is tied to the large $\pi\Sigma(1385)\to\pi\Sigma(1385)$
amplitude. This is why we consider this mechanism mostly
responsible for the strength in this region, particularly at
energies close to the tail of the $\Lambda(1520)$ resonance,
in spite that there are other terms of non resonant nature
that can produce a background there, which we are not
considering. 
We would also like to note that most models for the
$\Lambda(1520)$ (like quark models) that just provide a mass
and a width for the resonance, would lead to a distribution (in
the absence of the background terms neglected by us) given
approximately by the Breit-Wigner distribution shown in the
figure. The difference between this Breit-Wigner form and the
distribution of our model is a genuine consequence of the
unitary chiral dynamics assumed in our approach.

\begin{figure*}[htb]
\begin{center}
\includegraphics[height=9cm]{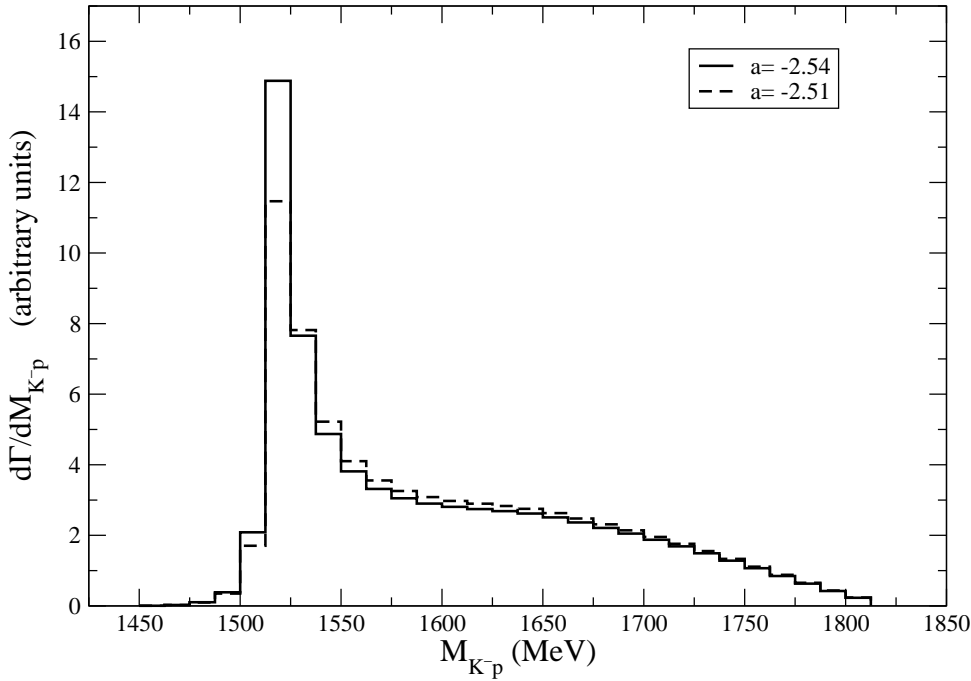}
\end{center}
\caption{Same as Fig.~\ref{fig:dares} but 
for $E_\gamma=2.0-2.4\textrm{ GeV}$
and implementing a
binning with the experimental resolution of $12.5\MeV$.} 
\label{fig:spring8}
\end{figure*}

It is interesting to see what our model gives for
$d\sigma/dM_{K^-p}$ at the photon energies of
Spring8/Osaka $E_\gamma=2.0-2.4\textrm{ GeV}$. We show the
results in Fig.~\ref{fig:spring8} where we have made the average for
various energies of the beam in this interval and binned the
results with a resolution of $12.5\MeV$ of the experiment
\cite{nakano2}. We show the results for the two different
values of the subtraction constant which we used before to
account for uncertainties. We see from the results that there
is also a sizeable background at energies above the peak.
The agreement with the preliminary data of \cite{nakano2},
not shown in the figure, is
rather fair. This exercise has served to show consistency of
the preliminary data obtained in Spring8/Osaka with the old
data of Daresbury \cite{dares}, particularly concerning the
background beyond the $\Lambda(1520)$ peak, which is a matter
of concern when testing the ability of experimental methods to
deal with backgrounds and identify peaks on top of them.\\

The other issue we address now is the evaluation of the cross
section for 
\begin{equation}  
  \gamma p \to K^+  \pi \Sigma(1385).
\label{eq:kk1}
\end{equation}

\begin{figure*}[htb]
\begin{center}
\includegraphics[height=1.8cm]{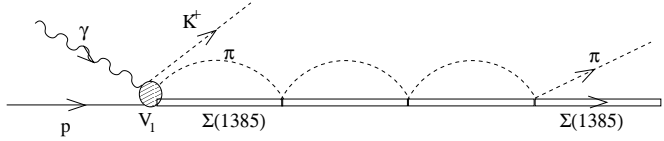}
\end{center}
\caption{Same as Fig.~\ref{fig:unit} for the 
$\gamma p\to K^+\pi\Sigma(1385)$ reaction.} 
\label{fig:unitb}
\end{figure*}

With our assumption that the  $\Lambda(1520)$ is a
dynamically generated resonance, mostly from $\pi
\Sigma(1385)$ interaction, the mechanism for the reaction of
Eq.~(\ref{eq:kk1}) is given in Fig.~\ref{fig:unitb},
 where we can see that it
shares with that of the $\gamma p \to K^+ K^- p$ reaction the
primary production of $\pi \Sigma(1385)$ and the full  $\pi
\Sigma(1385) \to \pi \Sigma(1385)$ scattering matrix. 
 Only the final vertices, leading to different final states,
are different between these reactions. We can easily
establish a link between these two vertices using
simultaneously empirical information of the $\Lambda(1520)$
decay into  $K^- p$ and the theoretical value for the
coupling of the $\Lambda(1520)$ to $\pi \Sigma(1385)$.

\begin{figure*}[htb]
\begin{center}
\includegraphics[height=1.8cm]{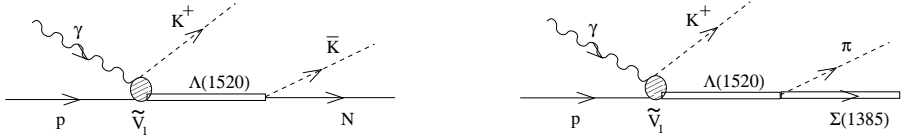}
\end{center}
\caption{Schematic representation of of $\gamma p \to K^+\bar
K N$ and $\gamma p\to K^+\pi\Sigma(1385)$ showing an explicit
$\Lambda(1520)$ exchange.} 
\label{fig:unitc}
\end{figure*}

In order to make the comparison between the two processes
clearer, we draw schematically in Fig.~\ref{fig:unitc} the
previous figures \ref{fig:unit} and \ref{fig:unitb},    

\noindent
which shows more transparently that what we need is the
coupling of the  $\Lambda(1520)$ resonance to the two decay
channels.  The coupling to the  $\pi \Sigma(1385)$ we already
found theoretically, see Eq.~(\ref{eq:pole}). The one to the
$K^- p$ we find now empirically. 
 From the PDG we know that the
partial decay width of the  $\Lambda(1520)$ to $\bar{K} N$ is
7.02 MeV.  Since the vertex is of D-wave type we can take
for it 
\begin{equation} 
\frac{h}{m_K^2} q^2,
\end{equation}
with the
same normalization as the coupling of the $\Lambda(1520)$ to 
$\pi \Sigma(1385)$, by means of which the $\Lambda(1520)$
partial decay width into $\bar{K} N$ is given by

\begin{equation}
\Gamma_{\Lambda(1520)\to\bar{K} N}
=\frac{1}{2\pi}\left(\frac{h}{m_K^2}\right)^2
\frac{M_N}{M_\Lambda} q^5  \qquad ; \qquad (q=244\textrm{
MeV}).
\end{equation}
This gives $h=2.21$.

All this said, the ratio of cross sections for the two
reactions is given by

\begin{equation}
R=\frac {(d\sigma_{\gamma p\to
 K^+\pi\Sigma^*}^{I=0})/dM_{\pi\Sigma^*}}
 {(d\sigma_{\gamma p\to K^+\bar K N}^{I=0})
/dM_{\bar K
N}}
 =\frac{g^2}{(h/m_K^2)^2}
  \frac{\int dE'_1 M_I \Theta(1-A'^2)
  \,|t_{\pi\Sigma^*\to\pi\Sigma^*}|^2}
  {\int dE_1 M_I \Theta(1-A^2) q^4 
 \,|t_{\pi\Sigma^*\to\pi\Sigma^*}|^2}
 \label{eq:R}
 \end{equation}
with $A'$ the corresponding $A$ variable, Eq.~(\ref{eq:A}),
for the kinematics of the $\gamma p\to K^+\pi\Sigma(1385)$ 
channel, and $M_I$ is the common invariant mass of
 the $\pi^0\Sigma^{*0}$ and $K^-p$ system.
The $\pi \Sigma(1385)$ scattering matrix cancels in the
numerator and denominator in Eq.~(\ref{eq:R}).
  Thus the ratio of mass
distributions for the two processes  is given by the ratio of
the couplings squared and the phase space, including the
factor $q^4$ in the $\gamma p\to K^+\bar K N$ channel.

   So far we have not made any distinction for the charge of
the final states since we have being using the isospin basis
and have taken the  $\pi \Sigma(1385)$ amplitudes in $I=0$, the
channel of the $\Lambda(1520)$. The isospin decomposition of
the $\bar{K}N$ and $\pi \Sigma(1385)$ states is given in
Eqs.~(\ref{eq:iso1}) and (\ref{eq:iso2})

\begin{eqnarray} \nonumber
    |K^-p\rangle &=& -\frac{1}{\sqrt{2}}|1, 0\rangle +
     \frac{1}{\sqrt{2}}|0, 0\rangle \\
    |\bar K^0n\rangle &=&  \frac{1}{\sqrt{2}}|1, 0\rangle +
     \frac{1}{\sqrt{2}}|0, 0\rangle
  \label{eq:iso1}
\end{eqnarray}

\begin{eqnarray}
    |\pi^+\Sigma^{*-}\rangle &=& - \frac{1}{\sqrt{6}}|2, 0\rangle - \frac{1}{\sqrt{2}}|1, 0\rangle -
     \frac{1}{\sqrt{3}}|0, 0\rangle  \nonumber \\
    |\pi^-\Sigma^{*+}\rangle &=&  \frac{1}{\sqrt{6}}|2, 0\rangle
    - \frac{1}{\sqrt{2}}|1, 0\rangle +
     \frac{1}{\sqrt{3}}|0, 0\rangle  \nonumber \\
|\pi^0\Sigma^{*0}\rangle &=&  \sqrt{\frac{2}{3}}|2, 0\rangle -
\frac{1}{\sqrt{3}}|0, 0\rangle .
\label{eq:iso2}
\end{eqnarray}
  
  We see in the equations that, neglecting the $I=2$, for which
the amplitudes in the chiral unitary approach are very small,
the $\pi^0 \Sigma^0(1385)$ channel is purely of $I=0$.  This
channel is hence ideal to isolate the $I=0$ term.  On the
other hand, similarly to what was done in \cite{nacher} in
the photoproduction of the $\Lambda(1405)$, one can see that
the cross sections are proportional to 

\begin{eqnarray}
\frac{1}{2}|T^{(1)}|^2 + \frac{1}{3}|T^{(0)}|^2 
+ \frac{2}{\sqrt{6}}Re(T^{(0)} T^{(1)*})\,
 ;\hspace{0.5cm} \pi^+\Sigma^{*-} \nonumber \\
\frac{1}{2}|T^{(1)}|^2 + \frac{1}{3}|T^{(0)}|^2 
- \frac{2}{\sqrt{6}}Re(T^{(0)} T^{(1)*})\, 
;\hspace{0.5cm} \pi^-\Sigma^{*+} \nonumber \\
\frac{1}{3}|T^{(0)}|^2\, ;\hspace{0.5cm} \pi^0\Sigma^{*0}
\end{eqnarray}

Thus,  both the $\pi^0 \Sigma^0(1385)$ channel and the
average of the  $\pi^+ \Sigma^-(1385)$ and $\pi^-
\Sigma^+(1385)$ cross sections can be used to isolate the
$I=0$ cross section, provided that the $I=1$ cross section is
relatively small compared to the $I=0$ one.  This exercise of
removing the interference term between $I=0$ and $I=1$, may
be important in case the $I=1$ component were not much
smaller that the $I=0$ one, since $|T^{(1)}|^2$ can be small
compared to $|T^{(0)}|^2$ but not the interference term. 
This could be the case since in the region close to
$1670\MeV$, there is another $3/2^-$ dynamically generated
resonance, the $\Sigma(1670)$,  constructed from the same
building blocks than  the  $\Lambda(1520)$, $\pi
\Sigma(1385)$, together with $\Delta \bar{K}$ to which  the
resonance couples with largest strength.  Actually, the
$\Sigma(1670)$ should in principle be  already seen in the
experiment of Spring8, \cite{nakano2}, but there is no trace
of this resonance in this experiment. This could be
understood in terms of the small branching ratio of that
resonance to the $\bar{K}N$ system, which is  about $10$
percent according to the PDG by contrast to the $45$ percent
of the  $\Lambda(1520)$.  On the other hand, we also have a
small branching ratio of the $\Sigma(1670)$ to $\pi
\Sigma(1385)$ which is also of the order of 10 percent as
reconstructed from the $(\Gamma_1 \Gamma_7)^{1/2}/\Gamma$
ratio of the PDG. This indicates that the background of $I=1$
in the spectrum of $\bar{K}N$, or in the one of $\pi
\Sigma(1385)$, should be relatively small, and we can rely upon
the $I=0$ dominance of the amplitudes of Eq.~(\ref{eq:t}),
particularly in the region between the $\Lambda(1520)$ and
$\Sigma(1670)$,  this is, around 1550-1630 MeV. Given the
Clebsch-Gordan coefficients for the isospin decomposition of
the  $K^- p$ and $\pi^0 \Sigma^0(1385)$ states in 
Eqs.~(\ref{eq:iso1}) and (\ref{eq:iso2}), we then conclude
that the ratio of mass distributions for these observable
channels is given by

\begin{equation}
R=\frac {(d\sigma_{\gamma p\to
 K^+\pi^0\Sigma^{*0}})/dM_{\pi^0\Sigma^{*0}}}
 {(d\sigma_{\gamma p\to K^+K^- p})
/dM_{K^-p}}
 =\frac{1/3}{1/2}\frac{g^2}{(h/m_K^2)^2}
 \frac{\int dE'_1 \Theta(1-A'^2)  }
 {\int dE_1  \Theta(1-A^2) q^4}
\label{eq:Rb}
\end{equation}

For the reasons given above, the average of the 
$\pi^+ \Sigma^-(1385)$ and $\pi^- \Sigma^+(1385)$ cross sections could be
similarly used instead of the $\pi^0 \Sigma^0(1385)$ one, which is not observable
at present in some labs like Spring8. 

\begin{figure*}[htb]
\begin{center}
\includegraphics[height=9cm]{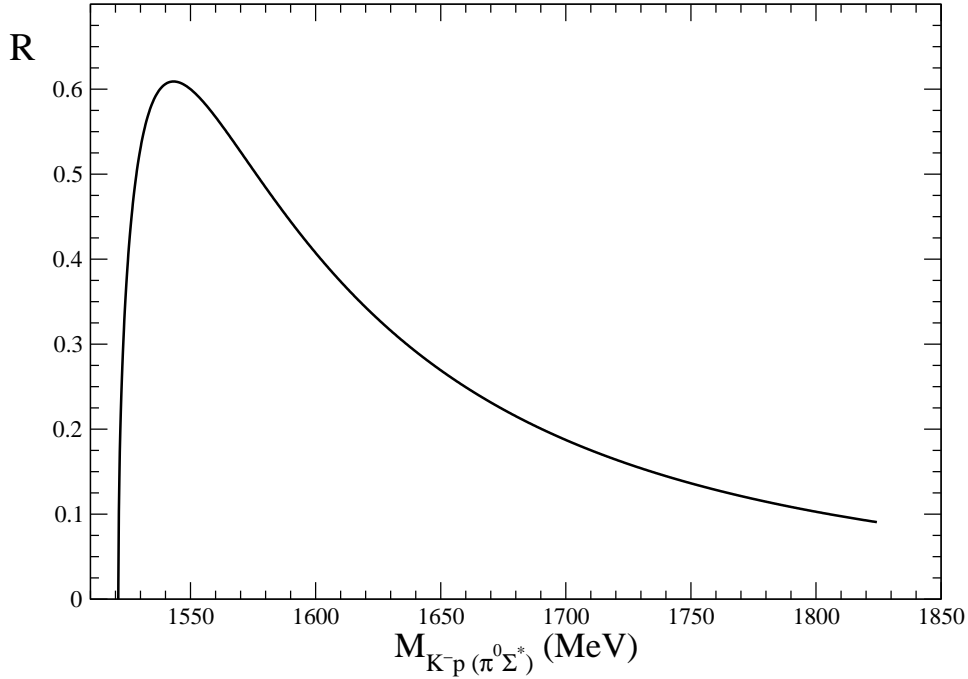}
\end{center}
\caption{Predicted ratio of the cross sections between the
$\gamma p\to
 K^+\pi^0\Sigma^{*0}$ and $\gamma p\to K^+K^- p$ reactions.
 (See Eq.~(\ref{eq:Rb})).} 
\label{fig:ratio}
\end{figure*}

In Fig.~\ref{fig:ratio} we plot the ratio of Eq.~(\ref{eq:Rb})
 as a function of the $K^- p$ and $\pi^0 \Sigma^0(1385)$ 
 invariant mass for $|g|=1.21$ and $E_\gamma=2.4\textrm{ GeV}$.
 In our model, this
ratio is independent of the photon energy. However, the photon
energy is relevant to establish the maximum invariant mass 
where the
ratio can be defined. We can see in Fig.~\ref{fig:ratio}
 that, in
the region we suggest, $1550-1630\MeV$, the ratio $R$ decreases
from values around $0.6$ to $0.3$. Note that the bump in
Fig.~\ref{fig:ratio} has nothing to do with the
 $\Lambda(1520)$
resonance, since the amplitudes producing it have canceled in
the ratio. It comes essentially from the phase space of the
two reactions. With assumed uncertainties of about $20$
percent in $|g|$, which would lead to about $40$ percent
uncertainties in the ratio of Fig.~\ref{fig:ratio}
 and extra
uncertainties from approximations done, we assume that an
error of about $50$ percent is a conservative estimate of the
uncertainties in the calculations.

The results obtained are essentially related to the
theoretical coupling, $g$, of the $\Lambda(1520)$ to the $\pi
\Sigma(1385)$ channel which is predicted by the
theory.  Hence the actual measurement of the ratio discussed,
when compared  with  theoretical predictions, would produce
an  experimental  measurement of that coupling  which could
substantiate the claim that the $\Lambda(1520)$ is a
dynamically generated resonance from the interaction of the
$\pi \Sigma(1385)$ and $K \Xi(1530)$ coupled channels, and
particularly from the first one. 

  At the same time it would be interesting to evaluate the same coupling with
present quark models to see if there are substantial differences, such that the
experimental determination of the coupling could clearly favor one picture over
the other one.\\

Summarizing the results, we have done a study of some
implications of the nature of the $\Lambda(1520)$ resonance
from a perspective of chiral unitary dynamics in which
framework the resonance appears as a dynamically generated
state, mostly from the $\pi\Sigma(1385)$ interaction in $L=0$
and $I=0$.

First we have addressed the origin of the cross section for the
$\gamma p\to K^+K^-p$ reaction at invariant masses close and
above the $\Lambda(1520)$ mass. The dynamical origin of the
$\Lambda(1520)$ within the chiral unitary approach, together
with the $D-$wave character of the $\Lambda(1520)\to \bar K N$
decay, are responsible in our model for this relatively large
strength, which is tied to a large coupling of the
$\Lambda(1520)$ to the $\pi\Sigma(1385)$ channel and a fairly
large $\pi\Sigma(1385)\to\pi\Sigma(1385)$ amplitude in $I=0$.

Second we have made predictions of the ratio of the 
$\gamma p\to K^+\pi\Sigma(1385)$ to the $\gamma p\to K^+K^-p$
cross sections, which is closely tied to the value of the
$\Lambda(1520)$ coupling to $\pi\Sigma(1385)$, and which is
provided by the chiral unitary approach. Therefore, an
experimental measurement of such ratio would provide a test of
the claimed nature of this resonance.

\section{Acknowledgments}
One of us, L.R., acknowledges support from the 
Ministerio de Educaci\'on y Ciencia.
We would like to thank T.~Nakano for
useful discussions.
This work is partly supported by the Spanish CSIC and JSPS collaboration, the
 DGICYT contract number BFM2003-00856,
and the E.U. EURIDICE network contract no. HPRN-CT-2002-00311.

%\appendix
%\section{}
%\label{ap1}


\begin{thebibliography}{99}

%\cite{Kaiser:1995eg}
\bibitem{norbert}
N.~Kaiser, P.~B.~Siegel and W.~Weise,
%``Chiral dynamics and the low-energy kaon - nucleon interaction,''
Nucl.\ Phys.\ A {\bf 594} (1995) 325
[arXiv:nucl-th/9505043].
%%CITATION = NUCL-TH 9505043;%%

%\cite{Kaiser:1995cy}
\bibitem{siegel}
N.~Kaiser, P.~B.~Siegel and W.~Weise,
%``Chiral dynamics and the S11 (1535) nucleon resonance,''
Phys.\ Lett.\ B {\bf 362} (1995) 23
[arXiv:nucl-th/9507036].
%%CITATION = NUCL-TH 9507036;%%

%\cite{Oset:1997it}
\bibitem{angels}
E.~Oset and A.~Ramos,
%``Non perturbative chiral approach to s-wave anti-K N interactions,''
Nucl.\ Phys.\ A {\bf 635} (1998) 99
[arXiv:nucl-th/9711022].
%%CITATION = NUCL-TH 9711022;%%

%\cite{Meissner:1999vr}
\bibitem{jose}
U.~G.~Meissner and J.~A.~Oller,
%``Chiral unitary meson baryon dynamics in the presence of resonances:  Elastic
%pion nucleon scattering,''
Nucl.\ Phys.\ A {\bf 673} (2000) 311
[arXiv:nucl-th/9912026].
%%CITATION = NUCL-TH 9912026;%%


%\cite{Meissner:1999vr}
\bibitem{ulf}
U.~G.~Meissner and J.~A.~Oller,
%``Chiral unitary meson baryon dynamics in the presence of resonances:  Elastic
%pion nucleon scattering,''
Nucl.\ Phys.\ A {\bf 673} (2000) 311
[arXiv:nucl-th/9912026].
%%CITATION = NUCL-TH 9912026;%%

%\cite{Oset:2001cn}
\bibitem{bennhold}
E.~Oset, A.~Ramos and C.~Bennhold,
%``Low lying S = -1 excited baryons and chiral symmetry,''
Phys.\ Lett.\ B {\bf 527} (2002) 99
[Erratum-ibid.\ B {\bf 530} (2002) 260]
[arXiv:nucl-th/0109006].
%%CITATION = NUCL-TH 0109006;%%

%\cite{Ramos:2002xh}
\bibitem{bennhold2}
A.~Ramos, E.~Oset and C.~Bennhold,
%``On the spin, parity and nature of the Xi(1620) resonance,''
Phys.\ Rev.\ Lett.\  {\bf 89} (2002) 252001
[arXiv:nucl-th/0204044].
%%CITATION = NUCL-TH 0204044;%%

%\cite{Jido:2003cb}
\bibitem{jido}
D.~Jido, J.~A.~Oller, E.~Oset, A.~Ramos and U.~G.~Meissner,
%``Chiral dynamics of the two Lambda(1405) states,''
Nucl.\ Phys.\ A {\bf 725} (2003) 181
[arXiv:nucl-th/0303062].
%%CITATION = NUCL-TH 0303062;%%

%\cite{Hyodo:2002pk}
\bibitem{nam}
T.~Hyodo, S.~I.~Nam, D.~Jido and A.~Hosaka,
%``Flavor SU(3) breaking effects in the chiral unitary model for meson baryon
%scatterings,''
Phys.\ Rev.\ C {\bf 68} (2003) 018201
[arXiv:nucl-th/0212026].
%%CITATION = NUCL-TH 0212026;%%

%\cite{Kolomeitsev:2003kt}
\bibitem{lutz}
E.~E.~Kolomeitsev and M.~F.~M.~Lutz,
%``On baryon resonances and chiral symmetry,''
Phys.\ Lett.\ B {\bf 585} (2004) 243
[arXiv:nucl-th/0305101].
%%CITATION = NUCL-TH 0305101;%%

%\cite{Garcia-Recio:2003ks}
\bibitem{juan}
C.~Garcia-Recio, M.~F.~M.~Lutz and J.~Nieves,
%``Quark mass dependence of s-wave baryon resonances,''
Phys.\ Lett.\ B {\bf 582} (2004) 49
[arXiv:nucl-th/0305100].
%%CITATION = NUCL-TH 0305100;%%

\bibitem{pdg} 
S.~Eidelman {\it et al.}  [Particle Data Group Collaboration],
%``Review of particle physics,''
Phys.\ Lett.\ B {\bf 592} (2004) 1.
%%CITATION = PHLTA,B592,1;%%

%\cite{Sarkar:2004jh}
\bibitem{sarkar}
S.~Sarkar, E.~Oset and M.~J.~Vicente Vacas,
%``Baryonic resonances from baryon decuplet - meson octet interaction,''
arXiv:nucl-th/0407025.
%%CITATION = NUCL-TH 0407025;%%

%\cite{Sarkar:2004sc}
\bibitem{kdel}
S.~Sarkar, E.~Oset and M.~J.~Vicente Vacas,
%``A resonant Delta K state as a dynamically generated exotic baryon,''
arXiv:nucl-th/0404023.
%%CITATION = NUCL-TH 0404023;%%

%\cite{Nakano:2003qx}
\bibitem{nakano}
T.~Nakano {\it et al.}  [LEPS Collaboration],
%``Observation of S = +1 baryon resonance in photo-production from  neutron,''
Phys.\ Rev.\ Lett.\  {\bf 91} (2003) 012002
[arXiv:hep-ex/0301020].
%%CITATION = HEP-EX 0301020;%%


\bibitem{pentawork} http://www.rcnp.osaka-u.ac.jp/~penta04/

\bibitem{nakano2} T. Nakano, talk at the Pentaquark04 workshop \cite{pentawork}.

\bibitem{dares}
D.~P.~Barber {\it et al.},
%``Strangeness Exchange In The Photoproduction Of K+ Lambda (1520) Between
%2.8-Gev And 4.8-Gev,''
Z.\ Phys.\ C {\bf 7} (1980) 17.
%%CITATION = ZEPYA,C7,17;%%

\bibitem{chan}
S.~B.~Chan, J.~Button-Shafer, S.~S.~Hertzbach, R.~R.~Kofler and M.~Schiff,
%``Study Of Sigma(1385) Pi Production Near The Lambda(1520) Resonance,''
Phys.\ Rev.\ Lett.\  {\bf 28} (1972) 256.
%%CITATION = PRLTA,28,256;%%
%\cite{Nacher:1998mi}

\bibitem{nacher}
J.~C.~Nacher, E.~Oset, H.~Toki and A.~Ramos,
%``Photoproduction of the Lambda(1405) on the proton and nuclei,''
Phys.\ Lett.\ B {\bf 455} (1999) 55
[arXiv:nucl-th/9812055].
%%CITATION = NUCL-TH 9812055;%%








\end{thebibliography}
\end{document}